\documentclass[reprint, amsmath, amssymb, preprintnumbers, showpacs,
  showkeys, aps, prb]{revtex4-1}
\usepackage{graphicx}
\usepackage{color}
\usepackage{float}
\usepackage{nicefrac}
\begin{document}

\preprint{Beleanu {\it et al}}

\title{Large resistivity change and phase transition in LiMnAs}

\author{A. Beleanu}
\author{J. Kiss}
\author{G. Kreiner}
\author{C. K\"ohler}
\author{L. M\"uchler}
\author{W. Schnelle}
\author{U. Burkhardt}
\author{S. Chadov}
\author{S. Medvediev}
\author{D. Ebke}
\author{C. Felser}
\email{felser@cpfs.mpg.de}

\affiliation{Max-Planck-Institut f\"ur Chemische Physik fester Stoffe, Dresden, Germany}

\author{G. Cordier, B. Albert}
\affiliation{Technical University of Darmstadt, Eduard-Zintl-Institute of Inorganic and Physical Chemistry, Darmstadt, Germany}

\author{A. Hoser}
\affiliation{Helmholtz Zentrum Berlin f\"ur Materialien und Energie GmbH, Berlin, Germany}

\author{F. Bernardi}
\affiliation{Departamento de Fisica, Instituto de Fisica, Universidade Federal do Rio Grande do Sul (UFRGS), Porto Alegre, RS, Brazil}

\author{T.I. Larkin, D. Pr\"opper, A.V. Boris, B. Keimer}
\affiliation{Max-Planck-Institut f\"ur Festk\"orperforschung, Heisenbergstr. 1, D-70569 Stuttgart, Germany}

\date{\today}

\begin{abstract}
Antiferromagnetic semiconductors are new alternative materials for
spintronic applications and spin valves. In this work, we report a
detailed investigation of two antiferromagnetic semiconductors $A$MnAs
($A$\,=\,Li,~LaO), which are isostructural to the well known 
LiFeAs and LaOFeAs superconductors.  
Here we present a comparison between the structural, magnetic, and electronic
properties  of LiMnAs,  LaOMnAs  and  related materials.  
Interestingly,  both LiMnAs and LaOMnAs show a variation in resistivity
with more than five orders of magnitude, 
making them particularly suitable for use in future electronic devices.
Neutron and X-ray diffraction measurements on LiMnAs show
a magnetic phase transition corresponding to the Ne\'el temperature of 373.8~K, 
and a structural transition from the tetragonal to the cubic phase at 768~K.
These experimental results are supported by density functional theory (DFT) 
calculations.
\end{abstract}



\maketitle

\section{Introduction}
\begin{figure}
\centering
\includegraphics[width=0.9\linewidth,clip]{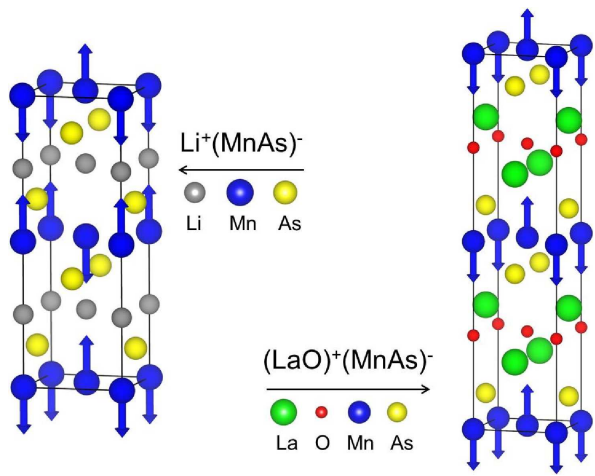}
\caption{(color online) Crystal and magnetic structure of LiMnAs~\cite{BMH86} and LaOMnAs\cite{EWS10}.
The blue arrows indicate the direction of the magnetic moments of the Mn atoms.
}
\label{fig:1}
\end{figure}

The idea of using semiconductors in spintronic devices has stimulated a new field 
of research. 
This is because by using not only the charge of the electron, but also its spin to
store and manipulate information has led to a new paradigm in condensed matter 
research to overcome the limitations set by Moore's law.\cite{ZFS04} 
Unfortunately, semiconductivity and ferromagnetism above room
temperature are not compatible.
Presently, to circumvent this issue, diluted magnetic semiconductors 
such as GaMnAs\cite{DOM00,JSM06} and LiZnMnAs\cite{scribd2,MKJ07} are used.
Recently, the use of antiferromagnetic (AFM) semiconductors has been proposed,
because they have a higher magnetic ordering temperature compared to diluted
magnetic semiconductors.
One promising candidate material for new spintronic applications is LiMnAs, 
given its suitable band gap and high magnetic moment per Mn atom.\cite{JNM11}
It was demonstrated that epitaxially grown LiMnAs thin films also show 
semiconducting behavior,\cite{JNM11} and their AFM ordering has been
confirmed\cite{JNM11,NCS11,WMH12} as well. 
Thus, LiMnAs, LaOMnAs, and related materials allow for the study and development of new 
AFM semiconductors, and the compounds are also interesting from a purely academic
viewpoint. 
CuMnAs\cite{MMS12} has also been suggested as an alternative semiconducting AFM
material; however, this compound shows semimetallic behavior in contrast to the 
desired semiconducting properties.
Still, the compounds with the general formula $A$Mn$X$, where $A$ = Li, Ni, Cu,
or LaO and $X$ = As or Sb offer a good starting point for the development of novel
semiconducting magnetic materials for applications in devices such as single-electron 
transistors.\cite{WJK06}\\
\indent
In the ground state, both LiMnAs and LaOMnAs show AFM insulating behavior. 
Chemically, LiMnAs consists of ionic Li$^{+}$ and tetrahedral [MnAs]$^{-}$
layers (see Fig.~\ref{fig:1}), with Mn$^{2+}$ in a $d^5$ and As$^{3-}$ 
in a nominal $s^{2}p^{6}$ closed shell configuration.
LiMnAs is isostructural to LiFeAs superconductor.\cite{Tapp_Tang2008}
Structurally, LaOMnAs is closely related to LiMnAs (see Figure~\ref{fig:1})
and to the LaOFeAs layered superconductor.
In comparison to LiMnAs, where the Mn moments are ordered antiferromagnetically
both in the $ab$ Mn planes and between the Mn planes stacked along the $c$ 
axis,\cite{BMH86} in the case of LaOMnAs, the moments are aligned
antiferromagnetically in the $ab$ plane but ferromagnetically along 
$c$.\cite{EWS10}\\
\indent
Interestingly, in contrast to the expected $5~\mu_{\rm B}$ of the 
${S=5/2}$ Mn$^{2+}$ ion, neutron scattering data on LiMnAs 
and LaOMnAs suggests a magnetic moment of approximately 3.4 to 3.8~$\mu_{\rm B}$,
which is close to the magnetic moment of 4~$\mu_{\rm B}$ for an \mbox{${S=2}$ Mn$^{3+}$} ion. 
This closely resembles the simple rules and general observation made 
regarding Heusler compounds~\cite{TCP2011}, stating that Mn tends
to have a large localized magnetic moment with an oxidation state of 3 
(in the literature referred to as the K\"ubler's rule\cite{KWS83}) 
in a nominal $d^4$ configuration. 
The magnetism then stems from the non-bonding
$d$ electrons, as can be best observed in Half-Heusler compounds, where only
Mn or rare-earth-containing compounds form stable magnetic
phases.\cite{KFS06,FFB07}\\
\indent
The aim of this work is multifold: we elaborate on the synthesis and bulk
characterization of polycrystalline LiMnAs and LaOMnAs, 
and compare their properties to other related compounds based on a short 
review from the literature on the $A$Mn$X$ family.
We show that LiMnAs can undergo a first-order phase transition from the
tetragonal to a cubic phase.
In addition, we discuss the electronic and magnetic properties of 
LiMnAs and LaOMnAs, including their transport properties.
By combining {\it ab-initio} density functional theory (DFT) calculations
with our experiments, we discuss the electronic structure
and magnetic exchange coupling of the Mn atoms, further commenting on the 
non-Curie-Weiss-like behavior experimentally reported for related 
compounds.\cite{SDM99}
We also address an open question arising from a previous
work~\cite{EWS10}, regarding the weak ferromagnetic behavior 
observed in LaOMnAs, which so far has been attributed to a small spin 
canting.
However, thus far, it was not taken into consideration that the formation of 
small amounts of MnAs as impurities during synthesis ($\leq${1\%}) 
cannot be excluded completely, and this could explain the weak 
ferromagnetic behavior.

\section{Results and discussion}
In Ref.~\onlinecite{MKS91} it was already noted, that in the $A$Mn$X$ series, 
where $A$ is an alkali metal, and $X$ is an element from the 15$^{th}$ group of the
periodic table, there is a correlation between the local moments of Mn and the
Mn--Mn distances ($d\rm_{Mn-Mn}$).
That is, a shorter $d\rm_{Mn-Mn}$ results in lower moment of Mn.
The $d\rm_{Mn-Mn}$ distance plays a major role in the magnetism of these
compounds, whose behavior is transferable to LiMnAs~\cite{BMH86}, LaOMnAs, and other materials.
The structural and magnetic data for the $A$MnAs series derived from neutron 
diffraction measurements is collected from the literature in Table~\ref{tab:Schuster}.
\begin{table}[t!]
\centering
\setlength\belowcaptionskip{0.5cm}
\setlength{\tabcolsep}{3.0pt}
\caption{
Short overview from the literature on the results obtained with neutron
diffraction measurements on $A$MnAs compounds, where $A$ is an alkali metal.
}
\begin{tabular}{lccccc}
  Compound            & $T$ [K]  & $a$ [{\AA}]& $c$ [{\AA}] & $d_{\rm
    Mn-Mn}$~[{\AA}] & $m$ [${\mu_{\rm B}}$]\\
   \hline
 LiMnAs$^*$               &  10    &  4.262(1)  &  12.338(1)  & 3.008(3) &     3.67(3)   \\
 NaMnAs~\cite{BMH86}  &  14    &  4.199(1)  &  14.164(1)  & 2.969(1) &     4.01(5)   \\ 
 KMnAs~\cite{BMH86}   &  14    &  4.382(1)  &  15.558(1)  & 3.099(1) &     4.03(5)   \\
 RbMnAs~\cite{SDM99}  &  10.6  &  4.400(5)  &  16.128(5)  & 3.111(1) &     4.05(4)   \\
 CsMnAs~\cite{MKS91}  &  10    &  4.417(1)  &  17.370(1)  & 3.123(1) &     4.07(2)   \\
 \hline
 \end{tabular}
 \\
\begin{flushleft}
*results from this work
\end{flushleft}
\label{tab:Schuster}
 \end{table}
%
Before turning to the presentation of our results, in Table~\ref{tab:IntroProp} 
first we provide a general overview of the most important characteristics 
and material properties of various compounds for the aforementioned $A$Mn$X$ class, with $A$ = Li, LaO.
Table~\ref{tab:IntroProp} serves as a set of reference values, providing 
a solid basis of comparison with our measured data for LiMnAs and LaOMnAs,
using which we can derive more general conclusions that are applicable to the
entire family.
As the data in Table~\ref{tab:IntroProp} show, the materials in the $A$Mn$X$ family can be
classified into two main categories: those that are semiconducting or
semi-metallic with AFM ordering, and those which are metallic or half-metallic 
and ferromagnetic (FM).
\begin{table*}[t!]
\setlength{\tabcolsep}{8.0pt}
\centering
\setlength\belowcaptionskip{0.5cm}
\caption{
Structural, electronic and magnetic properties of various $A$Mn$X$ based
compounds, where $A$ = Li, Ni, Cu, or LaO and $X$ = As or Sb, and the binary MnAs
compound. The shortest Mn--Mn distance is given by $d_{\rm Mn-Mn}$.
}
\begin{tabular}{lcccccc}
Compound &   Electronic  &  Magnetic & Mn magnetic        & $d_{\rm Mn-Mn}$ [{\AA}] & Space & Literature\\
         &   structure   &  order    & moment [${\mu_{\rm B}}$] &              & group   &\\
   \hline
  LiMnAs &  semiconductor & AFM      & 3.8                & 3.008     & {$P4/nmm$} & this work and Ref.~\onlinecite{BMH86}\\ 
         &                &          &                    &            & (tetragonal)      & \\[9pt] 
  CuMnAs & semimetal      & AFM      & no data                 & 3.452   & {$Pnma$} & Ref.~\onlinecite{MMS12}\\
         &                &          &                    &             & (orthorh.)      & \\[9pt] 
  CuMnSb & semimetal      & AFM      & 3.9                & 4.301       & {$F\bar{4}3m$} & Refs.~\onlinecite{MMS12,OFW87} \\
         &                &          &                    &             & (cubic)      & \\[9pt] 
 LaOMnAs & semiconductor  & AFM      & 3.34               & 2.914     & {$P4/nmm$}  & Refs.~\onlinecite{EWS10,EWS11}\\
         &                &          &                    &             & (tetragonal)      & \\[9pt] 
  NiMnAs & half metallic- & FM       & 4.0                & 4.051     & {$F\bar{4}3m$}  & Ref.~\onlinecite{DSK08}\\
         &  ferromagnet   &          &                    &             & (cubic)      & \\[9pt] 

  NiMnSb & half metallic- & FM       & 4.0                & 4.186       & {$F\bar{4}3m$} & Refs.~\onlinecite{EMW94,OFW87}\\
         &  ferromagnet   &          &                    &             & (cubic)      & \\[9pt] 

    MnAs & metallic       & FM       & 3.4                & 3.730      & {$P6_3/mmc$} & Refs.~\onlinecite{IKW06,HKY77}\\
         &                &          &                    &             & (hexagonal)      & \\
   \hline
 \end{tabular}
\label{tab:IntroProp}
 \end{table*}
There is a correlation between $d_{\rm Mn-Mn}$ distances, and the magnetic 
and electronic properties of these materials.
In other words, those compounds that have ${d_{\rm Mn-Mn}\le3.1}$~{\AA}
tend to be AFM semiconductors. In the range of ${3.1\le d_{\rm Mn-Mn}\le3.5}$~{\AA} 
they are semimetallic and AFM (except CuMnSb which has $d_{\rm Mn-Mn}$ of 4.301~${\AA}$ ), and those with 
${d_{\rm Mn-Mn}\ge3.5}$ are metallic and FM. 
As a general trend, this indicates that the further apart the Mn atoms are from each other, 
the more likely it is that they develop high moments, leading to a FM metal.
Hence, by adjusting the lattice parameters, the magnetic and electronic 
properties of the $A$Mn$X$ compounds can be tuned over a wide range.
Indeed, as an example, for LaOMnP (isostructural to LaOFeAs),
it was shown very recently that upon doping and external pressure, the system 
can undergo a phase transition from an AFM insulator to a metallic antiferromagnet.\cite{SYP12}

\subsection{Synthesis}
\indent
LiMnAs was synthesized and characterized as bulk material in the 
1980s~\cite{BMH86,AS81} by a high-temperature reaction of the
elements lithium, manganese, and arsenic.
It was found that this compound is AFM-ordered, where the N\'eel 
temperature is approximately 400~K, but the exact N\'eel 
temperature is not known.\\
\indent
Using a similar synthesis route, our preliminary experiments revealed a number 
of small Bragg reflections in powder X-ray diffraction (XRD) patterns of 
the target compound caused by impurities. 
These Bragg peaks become significantly stronger with increasing temperature and reaction 
time and stem from an attack of the tantalum ampoule by arsenic. 
Therefore, LiMnAs was prepared as follows: mixtures of stoichiometric amounts 
of Li (foil), Mn (chips) and As (pieces) (all 99.999\% purity)
were placed in Al$_{2}$O$_{3}$ crucibles. 
The total mass was approximately 2~g per sample. 
Each crucible was enclosed in an arc-sealed tantalum ampoule at 300~mbar Ar, 
which in turn was jacketed by an evacuated fused silica ampoule.
A grey powder was obtained after heating the mixtures to 1423~K with 
a low heating rate of 1~K/min in a muffle furnace.
This temperature was maintained for 2~h and then reduced
to 1173~K. After 24~h, the ampoules were quenched in water. 
All of the handling was performed in a glove-box under Ar
($p$(O$_2$,H$_2$O)$\leq1$~ppm).
LaOMnAs was synthesized from the starting materials LaAs, Mn, and MnO$_{2}$ 
in stoichiometric amounts at 1423~K for 42~h in an Al$_{2}$O$_{3}$ crucible, 
which was in turn enclosed in an evacuated and fused silica tube. 

\subsection{Crystal and magnetic structure}
\begin{figure}
\centering
\includegraphics[width=0.9\linewidth,clip]{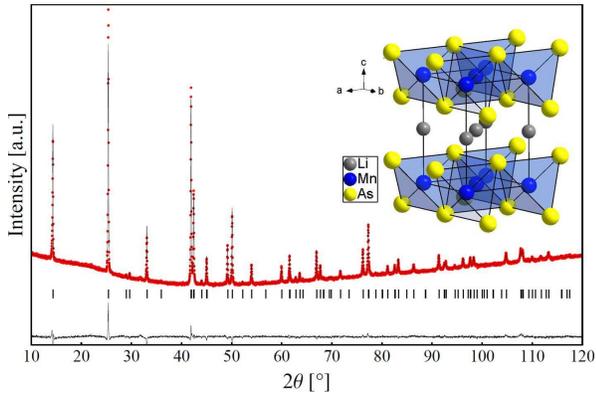}
\caption{(color online) The observed X-ray powder diffraction pattern (points) and the fit from
the Rietveld refinement (solid line) of LiMnAs at 293~K. 
The Bragg reflection marker and the difference curve are shown at the bottom. 
The inset shows the tetragonal crystal structure highlighting slabs of 
edge-connected MnAs$_4$ tetrahedra stacked along the $c$ axis.}
\label{fig:2}
\end{figure}

\begin{figure}
\centering
\includegraphics[width=1.0\linewidth,clip]{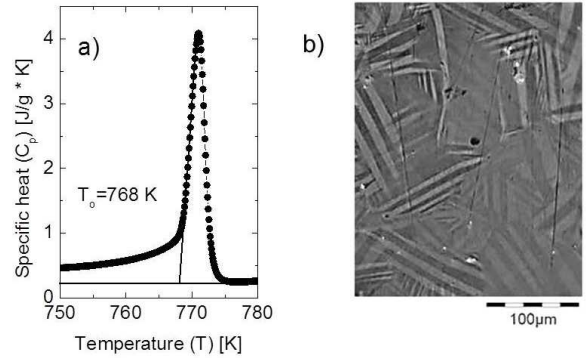}
\caption{(a) The DSC peak at 771~K with an onset temperature of ${T_{\rm o}=768}$~K
corresponds to the phase transformation to the cubic high-temperature phase.
(b) Optical bright field image at room temperature showing the domain structure of LiMnAs 
induced by the phase transformation.}
\label{fig:3}
\end{figure}

\begin{figure}
\centering
\includegraphics[width=1.0\linewidth,clip]{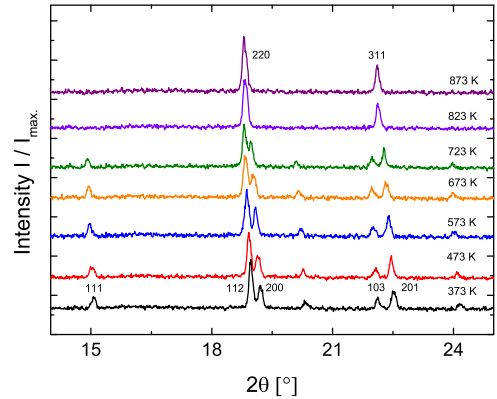}
\caption{(color online) Observed x-ray powder patterns in the range ${14^\circ\leq2\theta\leq25^\circ}$ 
of LiMnAs for various temperatures from 373~K to 873~K. The patterns up to 723~K
can be indexed primitive tetragonal. At 823~K and above, LiMnAs crystallizes in
a face-centered cubic structure.}
\label{fig:4}
\end{figure}

\begin{figure}
\centering
\includegraphics[width=0.8\linewidth,clip]{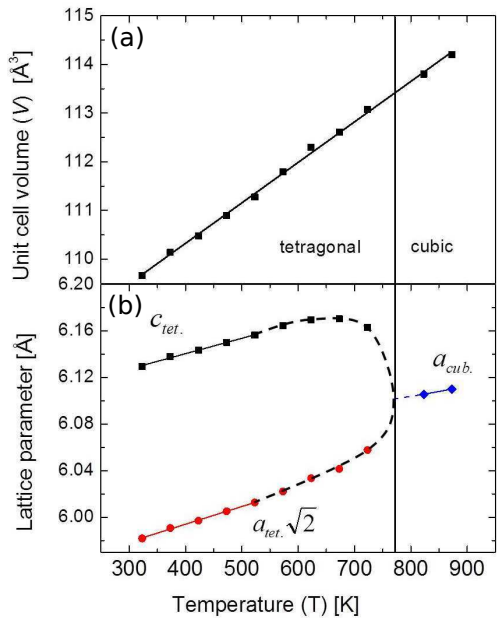}
\caption{(color online) (a) The unit cell volume of LiMnAs ($Z=4$) increases linearly
from 373~K to 873~K. There is no perceptible volume jump at the phase transition temperature.
(b) Lattice parameter (from the XRD measurements on LiMnAs) versus temperature for the
tetragonal and cubic phases. 
Below 600~K, the curves are fitted by a linear function. 
The dashed line at a higher temperature serves as a guide to the eye.}
\label{fig:5}
\end{figure}

\begin{figure}
\centering
\includegraphics[width=1.0\linewidth,clip]{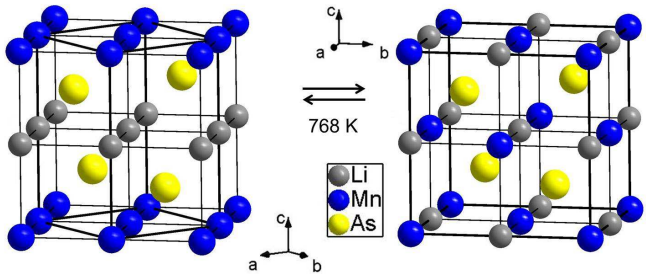}
\caption{(color online) Crystal structures of LiMnAs: (a)~tetragonal primitive and (b)~face-centered.}
\label{fig:6}
\end{figure}

\begin{figure}
\centering
\includegraphics[width=0.9\linewidth,clip]{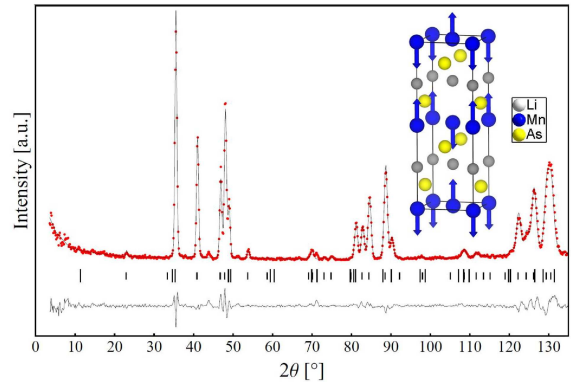}
\caption{(color online) Observed neutron powder pattern (points) for LiMnAs at 1.6~K with 
the fit of the Rietveld refinement of the nuclear and magnetic structure. 
The markers of the Bragg reflections for the nuclear reflections (short lines), 
magnetic reflections (long lines), along with the difference curve are shown at
the bottom. 
The inset shows the alignment of the magnetic moments along the $c$ axis.}
\label{fig:7}
\end{figure}

Powder X-ray diffraction (XRD) of the final product showed single-phase LiMnAs.
The powder diffraction experiment was performed with Cu K$\alpha_{1}$ 
radiation(${\lambda}$ = 1.5406~{\AA}) in Debye-Scherrer geometry. 
The sample was sealed in a glass capillary with a 0.3~mm diameter. 
A representative XRD pattern obtained at 293~K 
is shown in Figure~\ref{fig:2} with a Rietveld refinement carried 
out using Jana 2006 \cite{PDP06}. LiMnAs has a tetragonal unit cell with 
${a=4.2555(1)}$~{\AA} and ${c=6.1641(2)}$~{\AA} 
($P4/nmm$, ${Z=2}$, Li on $2b$, Mn on $2a$, 
As on $2c$ and $z$(As) = 0.7575(2)) in agreement with the older reports.\\
\indent
Chemical analyses for non-metal impurities were carried out using the carrier
gas hot extraction or the combustion technique (TCH 600, C 200 (LECO$^{\textregistered}$)).
For all samples, the impurities were below the limit of detection (LOD in ppm): 
H(100), N(200), O(1000) and C(1500) based on 20 mg initial weight. 
The 1:1:1 target composition was checked by Inductively-Coupled-Plasma-Optical Emission Spectrometry
 (ICP-OES; Vista RL, Varian) resulting in Li:Mn:As = 1.03(1):0.98(1):0.99(1).\\
\indent
A specimen for metallographic examination was prepared from one sample.
Because of moisture and air sensitivity, this was carried out in a glove-box using 
paraffin oil as lubricant. A bright field image of the microstructure is 
shown in Figure~\ref{fig:3}b, which reveals a domain structure. 
This is a typical signature for a tetragonal phase formed from a 
high-temperature cubic phase.\\
\indent
Specific heat measurements of LiMnAs in the temperature range of
240--780~K revealed only one DSC peak (see Figure~\ref{fig:3}a) 
corresponding to a reversible phase transition with onset temperature 
${T_{\rm o}=768}$~K.
To identify the nature of this phase transition, temperature-dependent
powder XRD was performed with a Stoe STADI II diffractometer with Mo K$\alpha_{1}$ radiation. 
The powder (20--50 $\mu$m) was sealed in a 0.5~mm quartz glass capillary. 
The measurements were carried out in the range from 323~K to 873~K in steps of 50~K. 
The observed powder patterns are shown in the range of
${14^\circ\leq2\theta\leq25^\circ}$ in Figure~\ref{fig:4}. All powder patterns up to 723~K can be indexed 
based on a tetragonal unit cell, whereas the powder patterns at 823~K and 873~K
are indexed based on a cubic unit cell with ${a_{\rm cub}\approx c_{\rm tet}\approx a_{\rm tet}\sqrt{2}}$.
The high-temperature phase, HT-LiMnAs, crystallizes in MgAgAs structure
type ($F\bar{4}3m$, ${Z=4}$, Li on $4b$, Mn on $4a$ and As on $4c$) according to the Rietveld analysis. 
The temperature dependence of the lattice parameters $a_{\rm cub}$, $c_{\rm
tet}$ and $a_{\rm tet}\sqrt{2}$ as well as the unit cell volume in a 
pseudo-cubic setting are plotted versus the temperature in Figure~\ref{fig:5}. 
Both parameters $a$ and $c$ increase nearly linearly with 
${\alpha_{a\sqrt{2}}=1.57\cdot10^{-4}\pm4\cdot10^{-6}}$~{\AA}/K and 
${\alpha_{c}=1.35\cdot10^{-3}\pm4\cdot10^{-6}}$~{\AA}/K up to 600~K. 
Above 650~K, $a$ increases and $c$ decreases until they match at 768~K, where ${a\approx6.101}$~{\AA}. 
The volume increases linearly with the temperature expansion coefficient
of ${\alpha_{\rm V}=0.00832\pm1\cdot10^{-6}}$~\AA$^3$/K.\\
\indent 
The structural relationship between tetragonal LiMnAs and the high-temperature cubic form
is shown in Figure~\ref{fig:6} with the tetragonal structure given in the pseudo-cubic setting. 
The crystal structure at room temperature form can be described as an alternate stacking of slabs 
composed of edge connected slightly distorted MnAs$_4$ and LiAs$_4$ tetrahedra. 
The MnAs$_4$ tetrahedra are slightly smaller than the LiAs$_4$ tetrahedra. 
At higher temperatures the $c/a$ ratio in pseudo-cubic setting 
approaches one. At 768~K, all tetrahedra become regular and half of the 
Li and Mn atoms of each layer exchange positions with each other in such a way 
that a Half-Heusler structure is obtained.
The diffusion of the Li atoms to the Mn sites and vice versa likely occurs
via the empty Li$_4$As$_4$ cubes, showing that Li is rather mobile in LiMnAs.
There is no perceptible discontinuity in the volume curve at the temperature
of the phase transition, and $c/a$ transits smoothly to one.
Although this fulfills the requirements for a second-order phase transition,
the fast exchange of Li and Mn cannot be described by a continuous process. 
This can therefore be only a first-order phase transition. 
The measured change in the entropy and the enthalpy at the phase transition 
temperature is ${\Delta{H}_{\rm trans}=2.19}$~J/(Kmol) (${\approx23}$~meV/f.u.) 
and ${\Delta{S}_{\rm trans}=2.85}$~J/(Kmol). 
For the magnetic characterization of LiMnAs, neutron powder diffraction patterns 
were recorded on the E6 Focusing Powder Diffractometer at Helmholtz-Zentrum Berlin, Germany,
using the wavelength ${\lambda=2.448}$~{\AA}. 
The measurements were performed at seven different temperatures in the range from
1.6~K to 873~K. Five diffraction patterns in the range of 1.6~K to 300~K correspond to  
tetragonal and antiferromagnetically ordered LiMnAs. 
To protect the sample against moisture and oxidation, a 10~mm fused silica tube 
was filled and sealed under vacuum. A representative neutron powder diffraction
pattern obtained at 1.6~K is shown in Figure~\ref{fig:7} along with the fit 
from the Rietveld refinement using the nuclear and magnetic structures. 
Table~\ref{tab:trans} lists crystallographic data obtained from the Rietveld refinements of
the crystal structures based on neutron powder patterns at various temperatures. 
The powder patterns below the ordering temperature reveal 
Bragg peaks of magnetic order that can be indexed with the modulation vector 
${\vec{k}=(0,0,\nicefrac{1}{2})}$. This corresponds to an antiferromagnetically ordered 
structure doubling the \textit{c}-axis of the nuclear structure. 
The refinement in the magnetic space group ${P_{C}4_{2}/ncm}$ (Belov notation) 
reveals that the magnetic moments at the Mn sites with $3.7~\mu_{\rm B}$ at 1.6~K 
are aligned parallel to the $c$ axis. These observations are in good
agreement with the previously reported data, except
that the magnetic space group was wrongly assigned in 
Ref.~\onlinecite{BMH86}. 
The low temperature magnetic moment of $3.7~\mu_{\rm B}$ is smaller than the 
value of $5~\mu_{\rm B}$ expected for the high-spin (${S=5/2}$) state of
Mn$^{2+}$, assuming a $g$-factor of 2 ($\mu=g\cdot S\cdot\mu_{\rm B}$).
\begin{table}[t]
\centering
\setlength{\tabcolsep}{5.0pt}
\caption{Crystallographic data for tetragonal antiferromagnetically ordered 
and tetragonal paramagnetic LiMnAs and for the high temperature cubic phase of LiMnAs. 
The $z$ parameter for As in the antiferromagnetically ordered structure 
is half of that of the paramagnetic structure. In a cubic setup it is
defined as ${1-z=\nicefrac{1}{4}}$. 
For comparison, the magnetic moment of Mn$^{2+}$ determined from neutron 
diffraction measurements is also given together with data from Ref.~\onlinecite{BMH86}.
The residual factors, $R_{\rm B}$, for the magnetic and nuclear parts are
not provided because they are nearly equal for each measurement (approximately 6\%).}

  \begin{tabular}{lccccc}
      $T$[K]            &   $a$ [{\AA}] &   $c$ [{\AA}]    &   V[\AA$^3$]
      & $z$   &    $m$[${\mu_{\rm B}}$]  \\
      \hline
   1.6                &  4.262(1)   &  12.338(1)  &  224.12	   	  &  0.3787(5)    &     3.72(3)   \\
   10                 &  4.262(1)   &  12.338(1)  &  224.12       &  0.3795(5)    &     3.68(3)   \\
   13~\cite{BMH86}    &  4.253(1)   &  12.310(1)  &  222.66       &  0.3800(1)    &     3.75(3)   \\
   30                 &  4.262(1)   &  12.338(1)  &  224.12       &  0.3796(5)    &     3.70(3)   \\ 
   50	                &  4.261(5)   &  12.337(9)  &  224.00       &  0.3789(1)    &     3.61(3)   \\
   293~\cite{BMH86}   &  4.267(1)   &  12.356(1)  &  224.97       &  0.3800(1)    &     2.59(2)   \\
   300                &  4.259(2)   &  12.333(4)  &  223.70       &  0.3813(1)    &     2.63(3)   \\
   317                &  4.269(2)   &  12.363(4)  &  225.20       &  0.3792(1)    &     2.40(3)   \\
   322                &  4.269(2)   &  12.366(4)  &  225.40       &  0.3796(1)    &     2.41(3)   \\
   332                &  4.270(2)   &  12.372(4)  &  225.60       &  0.3778(1)    &     2.25(3)   \\
   348                &  4.272(2)   &  12.374(4)  &  225.80       &  0.3771(1)    &     2.05(3)   \\
   352                &  4.273(2)   &  12.373(4)  &  225.80       &  0.3800(1)    &     1.90(3)   \\
   357                &  4.274(2)   &  12.377(4)  &  226.00       &  0.3783(1)    &     1.74(3)   \\
   362                &  4.273(2)   &  12.380(4)  &  226.00       &  0.3799(1)    &     1.61(3)   \\
   366                &  4.276(2)   &  12.380(3)  &  226.40       &  0.3802(4)    &     1.49(3)   \\
   370                &  4.275(3)   &  12.383(3)  &  226.20       &  0.3783(4)    &     1.15(3)   \\
   373                &  4.275(2)   &  12.382(3)  &  226.20       &  0.3814(3)    &     0.91(3)   \\
   393*~\cite{BMH86}   &  4.273(1)   &  12.370(1)  &  225.86       &  0.3800(1)    &     1.59(3)   \\
   423~\cite{BMH86}   &  4.279(1)   &   6.193(1)  &  113.39       &  0.7610(2)    &      -        \\
   573                &  4.289(5)   &   6.203(0)  &  114.11       &  0.7582(1)    &      -        \\
   873                &  6.155(4)   &      -      &  233.18       &  $\nicefrac{1}{4}$        &      -         \\
   \hline
  \end{tabular}
\\
\begin{flushleft}
*magnetic moment not compatible with our data
\end{flushleft}
\label{tab:trans}
  \end{table}

\begin{figure}
\includegraphics[width=1.00\linewidth,clip=]{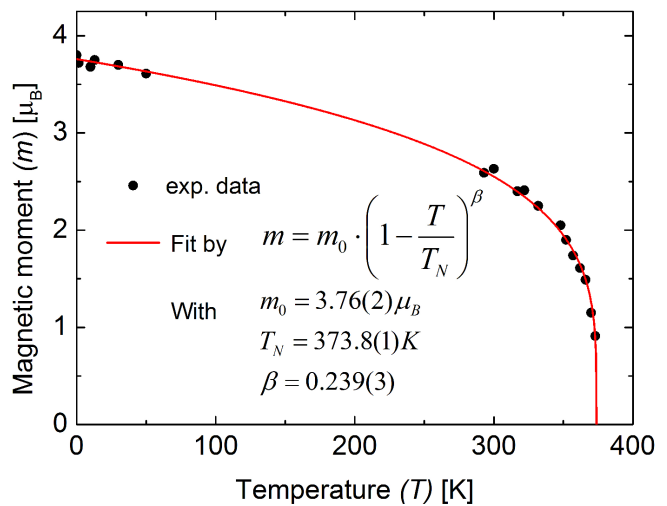}
\caption{Ordered Mn magnetic moment \textit{m} of LiMnAs as a function of 
temperature \textit{T}. 
The solid curve through the data is fitted to a power-law function
with critical exponent $\beta$.}
\label{fig:7a}
\end{figure}

\begin{figure}
\centering
\includegraphics[width=0.8\linewidth,clip]{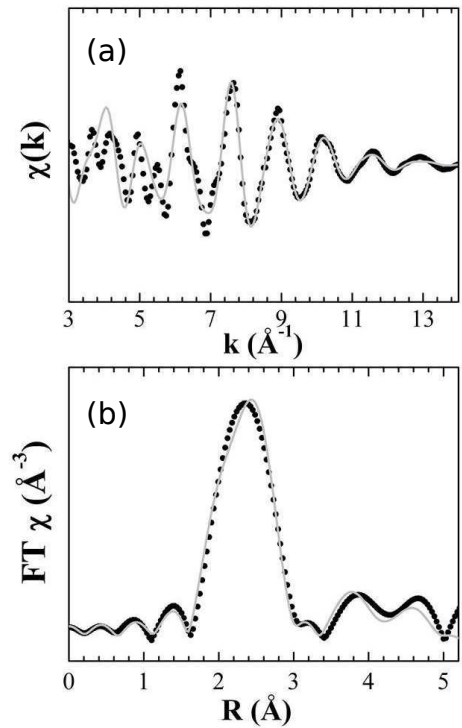}
\caption{(a)~293~K Mn EXAFS (points) and the corresponding best fit 
(solid line) of LiMnAs. (b)~Fourier transform of the 293~K Mn EXAFS (points) and 
the corresponding best fit (solid line) of LiMnAs.}
\label{fig:8}
\end{figure}

Figure~\ref{fig:7a} shows the temperature dependence of the magnetic moment 
\textit{m} obtained from the refinement of the neutron diffraction data.
All data below 380~K were fitted by the power law of ${m(T)} = {m_0}(1-T/T_{\textit{N}})^{\beta}$ 
with N$\acute{e}$el temperature, ${T_N}$ = 373.77~K and the critical exponent
$\beta$ = 0.239. 
Regarding the magnetism of LaOMnAs, in Ref.~\onlinecite{EWS10}, it was found
that it is antiferromagnetically ordered in the $ab$ plane, similar to 
LiMnAs, according to neutron diffraction analysis at lower temperatures.
The small increase in the magnetic moments was explained~\cite{EWS10} by a 
weak ferromagnetic behavior as a result of a small spin canting.
Such canting could be caused by broken inversion symmetry, giving rise to
the Dzyaloshinskii-Moriya~\cite{Dzyaloshinskii_58,Moriya_60} interaction 
modifying the magnetic ground state, which ceases to remain collinear.
However, in LiMnAs and LaOMnAs, the inversion symmetry break occurs anyway 
at their surfaces or interfaces. 
Thus, in bulk LaOMnAs we would expect, that canting plays a negligible role.
Instead of canting, we explain the small remanent total moment of these compounds 
with the formation of MnAs as an impurity phase.
This is because during synthesis, independent of the method used, 
tiny amounts of MnAs ($\leq1$\%) can easily be formed,
which cannot be detected by techniques such as XRD or energy-dispersive X-ray
spectroscopy (EDX).
As shown by the data in Table~\ref{tab:IntroProp}, MnAs is an ordered FM
material. It has a Curie temperature of 
approximately 320~K,\cite{FKK06} exactly coinciding with the magnetic transition temperature in our
measurements. Thus, the presence of MnAs impurities in the samples offer a more straightforward 
explanation for the small remanent total moments reported in
Ref.~\onlinecite{EWS10}, rather than spin canting.\\
\indent
Regarding the non-Curie-Weiss-like behavior of the magnetization as a function
of temperature reported for related $A$Mn$X$ compounds~\cite{SDM99},
our DFT calculations (see below) have shown that within the layers, the Mn
moments are very strongly coupled relative to the weak interlayer couplings.
Additionally, in LiMnAs between the layers, the Mn atoms are actually coupled
ferromagnetically with their next nearest neighbors  and antiferromagnetically
with their second nearest neighbors.
Because of this complex interaction between Mn moments, even above the N\'eel
temperature, a short-range local order is present, causing the 
non-Curie-Weiss-like behavior described above.
In comparison with theory, DFT calculations (see below) yield a value of 
$3.8~\mu_{\rm B}$ for the antiferromagnetically ordered structure at 0~K. 
In the refinement, the magnetic form factor for Mn$^{2+}$ has been used. 
Instead of Mn$^{2+}$, taking the magnetic form factor for Mn$^{+}$ 
results in a slight correlation with the magnetic moments, where the magnetic 
moment of Mn is enhanced by about 0.1-0.2~$\mu_{\rm B}$. 
However, in LiMnAs, referring to Mn$^{2+}$ makes chemically more logical. 
This is supported by analyzing the X-ray absorption near edge structure 
spectroscopy (XANES) data of several Mn compounds including LiMnAs.\\
\indent
The environment and the valence state of the Mn atoms in LiMnAs have been 
investigated by means of extended X-ray absorption fine-structure (EXAFS) 
and XANES at the Mn edge (6539 eV)~$K$ at room temperature in the 
transmission arrangement at the EXAFS beamline ${A1}$ 
of the Hamburg synchrotron radiation laboratory, HASYLAB, at DESY. 
Wavelength selection was realized by means of a Si(111) double-crystal 
monochromator. The near edge region of 6510-6560~eV was measured with a minimal 
step size of 0.25~eV, and the EXAFS regime was measured up to 7540~eV. 
A mass of approximately 10~mg of powdered Mn compounds with a particle size 
of $\leq20~\mu$m were diluted with $B_4C$ powder and mixed with 
polyethylene powder. 
The mixture was pressed to form a pellet with a diameter of 10~mm. 
All of the handling was performed under an argon atmosphere in a glove-box. 
Reference data were measured simultaneously with a Mn-foil.
This spectrum serves as an external reference for energy calibration and Mn valency. 
Figure~\ref{fig:8} shows the EXAFS signal, $\chi(k)$, and the 
corresponding Fourier transform. 
The EXAFS data were analyzed in accordance with the standard procedure for data 
reduction,\cite{KP88} using IFFEFIT.\cite{N01} 
The FEFF program was used to obtain the phase shift and amplitudes.\cite{ZRA95} 
The EXAFS signal $\chi(k)$ was extracted and Fourier-transformed (FT) 
using a Kaiser-Bessel window with a $\Delta(k)$ range of 7.0~\AA$^{-1}$. 
The $\chi(k)$ curve shows the characteristic pattern for this compound. 
The FT pattern shows one broad peak at approximately 2.2~{\AA} 
(uncorrected for the phase shift), 
which corresponds to the Mn--As and Mn--Mn scattering contributions. 
The Mn--Li scattering contribution is also included in this region; 
however, it does not have a strong influence on the main peak. 
The grey line in Figure~\ref{fig:8} represents the best fitting curve of the data. 
The structural parameters obtained from the Rietveld refinement were used as 
the input data to generate the cluster of atoms used. 
Only single scattering events were considered in the fitting procedure.  
The excellent agreement between the data and the theoretical structure was 
accomplished by the low $R$ factors($\leq0.6$\%).
 
A quantitative analysis extracted from the EXAFS data is summarized in Table~\ref{tab:trans2}. 
The amplitude-reduction term (S$_{0}$)$^{2}$ used was 0.75. 
The Mn--As and Mn--Mn distances are slightly contracted by $\leq2$\% and the 
Mn--Li by $\approx7$\% compared to the crystallographic data.

\begin{figure}
\centering
\includegraphics[width=0.9\linewidth,clip]{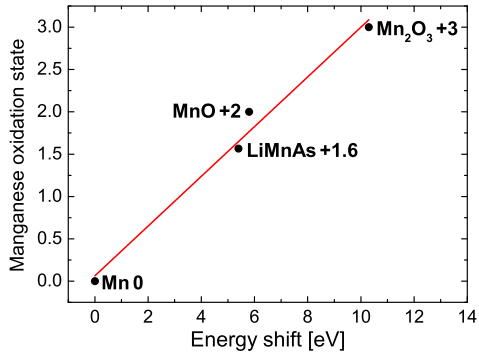}
\caption{Plot of the oxidation state of Mn in elemental manganese and in various
compounds versus the energy shift obtained from XANES measurements.}
\label{fig:9}
\end{figure}

\begin{table}[b]
\centering
\caption{Quantitative results for LiMnAs from the EXAFS data analysis for the 
Mn \textit{K} edge, considering the coordination numbers ($CN$) of the 
crystal structure. The best fitting exhibits the next-neighbor distances 
($R$), mean square displacement in $R$ $\sigma^2$, and $R$-factor for the whole fit.}
  \begin{tabular}{lccccccc}
      Shell  & Element  &   \textit{CN}  &  $d$ [{\AA}]  &  $R_{\rm
        EXAFS}$ [{\AA}]   &  $\sigma^{2}$ [{\AA}$^{2}$]     &  $R$ [\%]  \\
      \hline
   1st  &  As    &  4  &  2.601  &  2.550  &   0.006		 & \\
   2nd  &  Mn    &  4  &  3.012  &  2.953  &   0.006     & 0.6 \\
   3rd  &  Li    &  2  &  3.085  &  2.870  &   0.006    &   \\ 
\hline
   
  \end{tabular}
\label{tab:trans2}
  \end{table}
In Table~\ref{tab:trans2}, the distances obtained by the neutron-diffraction
refinement are compared with the EXAFS data.
The oxidation state of Mn in LiMnAs was obtained by using the 
XANES analysis with a widely used method as described by Wong {\it et al}.\cite{WLM84} 
This method considers the determination of the edge shift in relation to the 
Mn$^0$ edge energy for a given compound. 
In this case, there is a linear relation between this energy shift and the 
oxidation state. Using the oxidation state and energy shift of some standard 
oxides (Mn$_2$O$_3$ and MnO in this case), it is possible to find a linear 
relation, and obtain the oxidation state for the compound (Figure~\ref{fig:9}). 
This method reveals an oxidation state of Mn in 
LiMnAs of 1.6. Such a decrease in the oxidation state can be expected from 
weak homonuclear Mn--Mn interactions in the $ab$ plane.
This is supported by DFT calculations (see below), which show a considerable
interaction in the planes of Mn atoms, affecting their hybridization. 

\subsection{DC-resistivity}
\begin{figure}
\centering
\includegraphics[width=0.9\linewidth,clip]{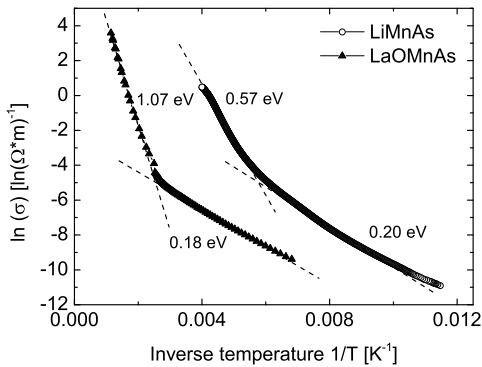}
\caption{Arrhenius plot of the electrical conductivity for LiMnAs and LaOMnAs.
At low temperatures, the curves correspond to the extrinsic regime, and at high
temperature, (low $1/T$), to the intrinsic regime.}
\label{fig:10}
\end{figure}

The resistivity of polycrystalline LiMnAs was recorded using the "Van der Pauw"
method in the temperature range of 50-300~K. 
For LaOMnAs, the electrical resistivities $\rho$(T) from 100-400~K were obtained 
by a standard linear four-point contact method. 
The measurement was performed by means of a physical property measurements system 
(PPMS, Quantum design model 6000, supported by LOT Germany). 
Bars of LaOMnAs $2\times2\times8$~mm$^3$ were cut from pellets obtained by a 
spark plasma sintering (SPS) process.\\
\indent 
Both LiMnAs and LaOMnAs show very high resistivity below ca.\@ 100~K.
The resistivity measurements suggest a semiconducting behavior. 
The Arrhenius equation and an Arrhenius plot were used in order to 
obtain the activation energy, which has a value of approximately
half of the band gap energy ${E{_g}}$ for intrinsic semiconductors~\cite{K73} according to
the relation:
\begin{equation}
\sigma_{\rm dc}\propto\exp(-\frac{E{_{\rm g}}}{2k{_{\rm B}}T}), 
\end{equation}
where ${k_{\rm B}}$ = Boltzmann constant (8.617332$\cdot10^{-5}$~eV/K), 
${T}$ is the absolute temperature, and ${E{_{\rm g}}}$ is the band gap width. 
Figure~\ref{fig:10} shows the Arrhenius plot and the temperature dependence of 
the conductivity measurements for LiMnAs and LaOMnAs. 
In the resistivity measurements both samples show the same behavior, namely, two distinct sections 
of activated conduction with a small ${E{_{\rm g}}}$ of 0.20~eV for LiMnAs and
an ${E{_{\rm g}}}$ of 0.18~eV for LaOMnAs, which indicate doped levels at lower temperatures and a larger 
${E{_{\rm g}}}$ of 0.57~eV for LiMnAs and an ${E{_{\rm g}}}$ of 1.07~eV for LaOMnAs respectively, 
at higher temperatures.
\begin{figure}
\centering
\includegraphics[width=0.9\linewidth,clip]{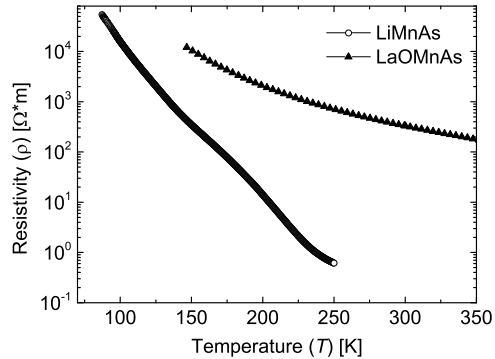}
\caption{Temperature dependence of the electrical resistivity, $\rho(T)$,
on a logarithmic scale for LiMnAs and LaOMnAs.}
\label{fig:11}
\end{figure}

Figure~\ref{fig:11} shows the resistivity as a function of temperature on a 
logarithmic scale. 
We note in particular, that the electrical resistivity changes by more than
five orders of magnitude for LiMnAs in the temperature range of 100--250~K 
and by about two orders of magnitude for LaOMnAs from 150--350~K. 

\subsection{Dielectric permittivity}

Ellipsometric measurements covering the far-infrared to deep-ultraviolet
spectral range (10~meV--6.5~eV) were carried out at room temperature 
on LaOMnAs ceramic dense pellets polished to an optical grade finish.
The measurements in the near-infrared to deep ultraviolet spectral range (0.75 eV - 6.5 eV) were performed 
with a rotating analyzer-type Woolam VASE variable-angle ellipsometer. 
For the infrared measurements from 0.01~eV to 1.0~eV, we used home-built ellipsometers in combination
with a Bruker Vertex 80v FT-IR spectrometer. 
Some of the experiments were performed at the infrared IR-1 beamline of the ANKA synchrotron light 
source at the Karlsruhe Institute of Technology, Germany.
The complex dielectric function, $\tilde \varepsilon(\omega) = \varepsilon_1(\omega)+\emph{i}\varepsilon_2(\omega)$ 
was directly determined from ellipsometric angles $\Psi(\omega)$ and $\Delta(\omega)$ \cite{Elli}. 
The  inversion of the ellipsometric data was performed within the framework of an effective medium approximation, which in the case of polycrystalline samples corresponds to the volume average of the anisotropic dielectric tensor projections. We did not take into account the surface roughness grain texturing effects on $\tilde \varepsilon(\omega)$, which is estimated to be less than 10$\%$ over the measured spectral range. 
Because of the air sensitivity, it was not possible to obtain ceramics of LiMnAs sufficiently dense for optical measurements.

\begin{figure}[ht]
\includegraphics[width=8.5cm]{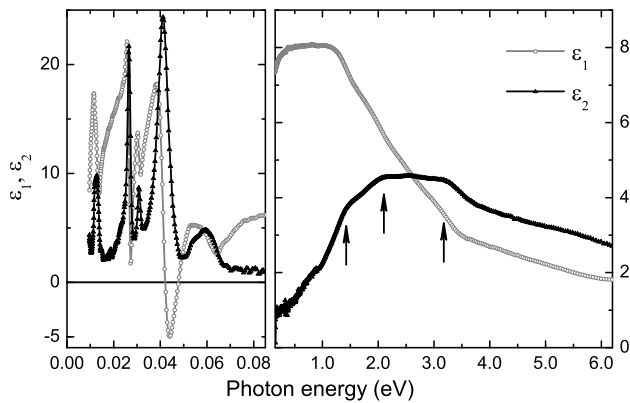} 
\caption{(Left) Phonon and (right) interband electronic contribution to the real and imaginary parts of the dielectric function $\tilde \varepsilon(\omega) = \varepsilon_1(\omega)+\emph{i}\varepsilon_2(\omega)$ of LaOMnAs measured at $T=294$ K. Arrows mark the peak positions of the main absorption bands.}
\protect\label{fig:12} 
\end{figure} 

Figure~\ref{fig:12} shows the complex dielectric function of the LaOMnAs sample at $T=294$ K. 
In the far-infrared range, there are six strong phonon modes. 
The lowest interband transition in this material lies above 0.4 eV. 
One can clearly distinguish three optical bands peaked at $\sim 1.4$, $\sim 2.1$, 
and $\sim 3.2$ eV.\cite{boris}. 
These transitions provide the main contribution to the static permittivity 
and form the direct absorption edge, below which the real part of the 
dielectric permittivity becomes nearly constant down to the phonon frequencies, 
$\varepsilon_1 (0.2 $ eV$ \lesssim \omega \lesssim 1.2$ eV$) \approx 8$. 
We note that the main features resemble those that characterize the interband 
electronic contribution to the dielectric function of polycrystalline LaOFeAs, 
while shifted by $\sim 0.7$ eV to higher energy \cite{boris}. 
The rise of $\varepsilon_2(\omega)$ above 0.4 eV can be attributed to indirect interband transitions. 

\subsection{DFT calculations}
\begin{figure}
\centering
\includegraphics[width=0.70\linewidth]{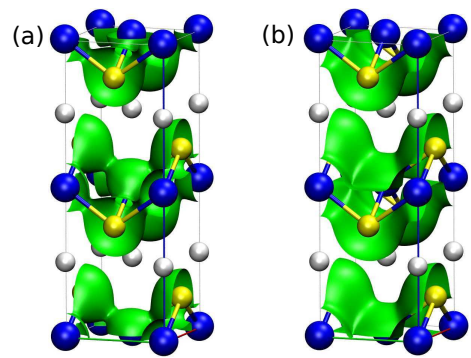}
\caption{(color online)
Real space electronic density of LiMnAs. The density
as green surfaces superimposed on top of the atomic structure 
at isovalues of $5.96\cdot10^{-3}$ and $3.58\cdot10^{-3}$~$e$/{\AA}$^3$,
corresponding (a)~to $sp^3$ hybridization  and (b)~to a mixture 
between Mn--As and Mn--Mn bonding.}
\label{fig:15}
\end{figure}

In order to gain further insights into the electronic structure and energetics
of LiMnAs and LaOMnAs, we have performed {\it ab-initio} calculations based on 
the density functional theory (DFT) with the VASP~\cite{VASP1,VASP2} code.
For the approximate treatment of the electron exchange and correlation, we have used the 
Perdew, Burke, Ernzerhof (PBE) functional~\cite{PBE} together with the 
projector-augmented wave (PAW) method with a plane-wave cutoff energy of 400~eV.
In the irreducible part of the Brillouin zone, the k-point integration was performed on an 
8$\times$8$\times$4 Monkhorst-Pack mesh.
To describe the magnetic ordering in the compounds as found from neutron
diffraction for the tetragonal antiferromagnetic structures, spin polarized calculations 
have been carried out for supercells containing four primitive unit cells 
(see Fig.~\ref{fig:1}).\\
\indent
In the case of LiMnAs, the calculated local Mn moments of 3.82~$\mu_{\rm B}$ for the
ground state are in excellent agreement with the value of 3.72~$\mu_{\rm B}$ 
measured at 1.6~K, which is the most representative temperature for a comparison between 
experiment and theory. 
As already mentioned in the introduction, the magnetic moment of the Mn
atoms is well localized in this class of materials. 
Based on a fully ionic picture of LiMnAs, i.\,e.\@ 
Li$^+$, Mn$^{2+}$, As$^{3-}$, one would expect that the oxidation state of Mn 
to be 2$^+$. 
The experimentally obtained value of 1.6$^+$ from XANES measurements
are somewhat lower than this 2$^+$ expected value.
Therefore, we have looked into the real-space density of LiMnAs, and 
apart from the expected $sp^3$ hybridization between Mn and As 
(see left panel in Fig.~\ref{fig:15}), we found that there is also
a feature present that corresponds to a mixture between Mn--Mn bonds 
and Mn--As bonds (see right panel in Fig.~\ref{fig:15}).

\begin{figure}
\centering
\includegraphics[width=0.80\linewidth]{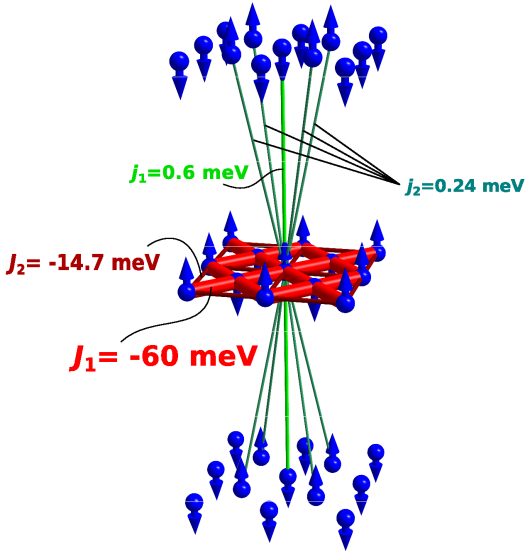}
\caption{(color online)
The scheme of the leading magnetic exchange interactions within the Mn atomic planes and
between the nearest Mn atomic planes of LiMnAs. For clarity, only Mn
atoms are shown (blue spheres) and the orientations of the magnetic moments 
are depicted with arrows. The thickness of the bonds connecting interacting
atoms are roughly proportional to the 
strength of the corresponding exchange interactions, where antiferromagnetic and ferromagnetic
coupling is represented by red ($J_{1,2}$) and green ($j_{1,2}$) colors, respectively.
}
\label{fig:CouplingConstants}
\end{figure}
From subsequent DFT calculations, we have determined the magnetic exchange coupling 
constants $J_{ij}$ of the classical Heisenberg model, 
${H=-\sum_{i>j}J_{ij}\hat{\rm e}_i\hat{\rm e}_j}$, where $\hat{\rm e}_{i,j}$
are the unit vectors corresponding to the directions of the local magnetization 
at sites $i$ and $j$.
For this purpose, we have employed the SPR-KKR program package,\cite{EKM11}
which is a suitable numerical tool for such computational task, not relying on
pseudopotentials, where $J_{ij}$ can be evaluated within the Green's function
formalism.\\
\indent 
As it follows from analysis of the exchange coupling scheme (see Fig.~\ref{fig:CouplingConstants}), 
within each Mn plane, the magnetic moments of the nearest neighbors are coupled
antiparallel (with ${J_1\approx-60}$~meV, shown in red), and this strong antiparallel
coupling cannot be perturbed by the competing but smaller next-nearest neighbor antiparallel
interaction (${J_2\approx-14.7}$~meV, shown in dark-red). 
Thus, within each plane the nearest Mn
magnetic moments are ordered antiferromagnetically. Further, to
understand how the nearest planes are magnetically oriented to one
another, we consider the inter-plane exchange coupling. Both
leading constants of the inter-plane exchange are positive,
${j_{1,2}\approx0.6}$ and 0.24~meV, which correspond to the nearest and
next-nearest inter-plane interactions, respectively. It is easy to see
that for an already fixed in-plane magnetic order, these two interactions are
also competing. Despite the fact that ${j_1>j_2}$, each Mn atom (for simplicity we
consider one in the center of Fig.~\ref{fig:CouplingConstants}) has only
2, but 8 next-nearest inter-plane neighbors, which leads to
an overall domination of the next-nearest inter-plane coupling
(${8j_2>2j_1}$); thus, the next-nearest Mn planes are coupled antiferromagnetically. \\
\indent
Based on the Mulliken analysis, the corresponding partial charges, i.\,e.\@, oxidation 
numbers of Li, Mn, and As were computed as $+0.8$, $+1.5$, and $-2.3$, respectively. 
This is again in good agreement with the XANES measurement presented
above, where an oxidation state of 1.6 was determined for the Mn atom in LiMnAs.
\begin{figure}
\centering
\includegraphics[width=0.8\linewidth,clip]{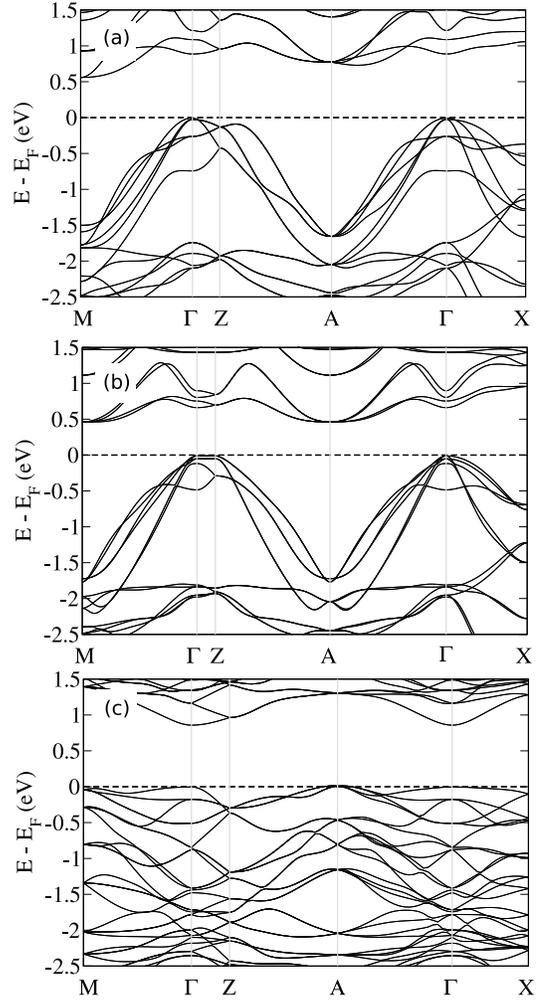}
\caption{Electronic band structure of tetragonal antiferromagnetic LiMnAs (a) and
LaOMnAs (b). In (c) the band structure of the cubic phase of LiMnAs is shown.
Spin-orbit coupling was not accounted.}
\label{fig:Bands}
\end{figure}
For LiMnAs, in our calculations we found an indirect band gap of 0.55~eV 
(see Fig.~\ref{fig:Bands}~(a)), which is very close to the 0.57~eV gap estimated from the higher 
temperature range of the Arrhenius plot of the conductivity measurements given in the previous
sections. 
These results are also in a good agreement with the recent full potential
calculations for bulk LiMnAs, where within the local density approximation (LDA) 
a $\Gamma$-M indirect gap of approximately 0.5~eV~\cite{JNM11} was computed.\\
\indent
As already shown, at 768~K LiMnAs undergoes a phase transition to the cubic crystal system.
Due to computational limitations and difficulties from the perspective of the
methodology, the implicit treatment of temperature effects in {\it ab-initio} 
calculations is not straightforward. 
Therefore, for the study of the cubic phase we have assumed, that the Mn atoms
are ordered antiferromagnetically as in the tetragonal phase. 
Using this model, our calculations indicate, that in contrast to the tetragonal
phase, the cubic structure has a direct band gap of 0.84~eV.
By comparing the computed total energies of the tetragonal and cubic phases,
we found that the cubic phase is only 24~meV higher in energy per formula unit
than the tetragonal phase. 
Thus, although the tetragonal structure is the ground state, the 
tetragonal-to-cubic phase transition can easily take place.
This is because the two phases are rather close in energy, and the energy 
difference of 24~meV is lower than the 75~meV thermal energy of the system at
768~K, which means that the system can easily surmount the energy difference
of 24~meV.
Hence, there is good agreement between the theoretically estimated
energy difference and the experimentally measured 23~meV/f.u.\@ 
change in enthalpy, which justifies the use of the 
simplified theoretical model.
Our calculations suggest that AFM ordering dominates in LiMnAs so strongly, 
to the extent that even if LiMnAs could be synthesized in the cubic phase, 
the pure undoped compound would still be an AFM semiconductor.\\
\indent
In the case of LaOMnAs, the theoretical local moments as obtained with the 
PBE functional for the Mn atoms are $3.54~\mu_{\rm B}$.
In contrast to these results, via neutron diffraction measurements performed at
290~K, the refined magnetic moment of Mn was $2.43~\mu_{\rm B}$, 
which is more than one $\mu_{\rm B}$ lower compared to our theoretical result. 
In a later experiment,\cite{EWS11} however, it was shown that the moments at 2~K 
are 3.34~$\mu_{\rm B}$, which is in good agreement with our calculations.
In the case of LiMnAs, our neutron-diffraction data also show a similar trend as a
function of temperature, where at 300~K the Mn moments are reduced from
3.72 to 2.62~$\mu_{\rm B}$ (see Table~1).
Furthermore, in case of the structurally closely related BaMn$_2$As$_2$ the Mn local moments
were determined\cite{SGH2009} to be 3.88~$\mu_{\rm B}$ at 10~K.
Hence, the magnetic properties and magnetic moments are rather similar in the
whole family of the $A$MnAs ($A$ = Li, LaO, Ba, etc.) layered structure type.\\
\indent
The computed oxidation state of the La, O, Mn, and As atoms are +2.4, -0.8, +1.4, and -3.1, 
respectively, i.\,e.\@, the oxidation state of O is underestimated, and that of As 
is overestimated.
Regarding the electronic structure, we found that LaOMnAs is also an indirect
band-gap semiconductor with an $E_{\rm g}$ of 0.46~eV (see Fig.~\ref{fig:Bands}~(b)),
which is considerably lower than the value of 1.07~eV derived from the Arrhenius plot.

\section{Conclusion}
In summary, in this paper, we have presented a broad overview of the synthesis,
atomic and electronic structure, and magnetic properties of LiMnAs, LaOMnAs, and other 
related compounds in the $A$Mn$X$ family.
Thus, with this work, we respond to several important questions raised concerning this 
important class of materials.
In general, the compounds in the $A$Mn$X$ family can be classified in two main categories: 
AFM semiconductors and FM metals.
The nature of the magnetic coupling and of the semiconducting or metallic 
behavior is mainly determined by the shortest Mn--Mn distances ($d\rm_{Mn-Mn}$), 
which seems to be a general property of the compounds containing MnAs layered networks.\\
\indent
For LiMnAs and LaOMnAs, our data are in good agreement with the literature,
showing that both are antiferromagnetic semiconductors, with a magnetic moment 
of ca. 4~$\mu_{\rm B}$ per Mn atom at low temperatures.
Interestingly, by combining thermal analysis and temperature dependent XRD
measurements, a phase transition to the cubic Half-Heusler phase has been discovered.
Theoretical calculations suggest that even if cubic LiMnAs could be stabilized,
the ideal undoped cubic counterpart would still be an AFM
semiconductor, similar to the tetragonal compound.
EXAFS and XANES measurements indicate the oxidation state of Mn to be 1.6. 
This is in agreement with the concept of localized moments on Mn and explains 
the close relationship between the cubic and tetragonal phase and the connection 
between Mn and rare earth ions in tetragonally coordinated structures. 
Regarding the weak ferromagnetism, in contrast to Refs.~\onlinecite{EWS10,EWS11}, 
due to symmetry considerations, we expect that canting takes place at the 
surfaces and interfaces of LiMnAs and LaOMnAs, but in the bulk it is unlikely.\\
\indent
Resistivity measurements on LiMnAs and LaOMnAs show two distinct regions of 
activated transport behavior, and a large change in resistivity by
more than five and two orders of magnitude, respectively. 
Furthermore, we have shown that LaOMnAs is also a promising material for the
realization of spin valves,\cite{PWM11} 
and our findings suggest that compounds such as BaMn$_2$As$_2$ and 
related compounds are promising candidates as well.

\bigskip

\section{Acknowledgments}

Financial support from the ERC Advanced Grant (291472) is gratefully acknowledged.
We are grateful to Susann Scharsach for the DSC measurements, 
Dr. Gudrun Auffermann for the chemical analysis measurements and 
Igor Veremchuk for the spark plasma sintering (SPS) measurements 
from the Max-Planck-Institute for Chemical Physics of Solids.  
Prof. F. Bernardi  received a research grant from CNPq - Brazil.
Financial support from the DFG project FOR 1464 “ASPIMATT” (1.2-A) 
is gratefully acknowledged. 
We also thank Y.-L. Mathis for providing support at the infrared 
beamline of the synchrotron facility, ANKA, at the Karlsruhe Institute of Technology.
 
\bibliography{beleanu}

\end{document}